%% file: probes_dnb.tex
\documentclass{mem}
\usepackage{natbib}
\usepackage{txfonts}
\usepackage{balance}
\usepackage{graphicx}
\usepackage[breaklinks,dvipdfm]{hyperref}
\idline{75}{282}

\begin{document}

\include{definitions}

\newcommand{\janus}{{\it JANUS}}

\title{Optimizing the Search for High-z GRBs:}
\subtitle{The JANUS X-ray Coded Aperture Telescope}

\author{
D.\,N.\,Burrows\inst{1,2} 
\and D.\,Fox\inst{1}
\and D.\,Palmer\inst{3}
\and P.\,Romano\inst{4}
\and V.\,Mangano\inst{4}
\and V.\,La\,Parola\inst{4}
\and A.\,D.\,Falcone\inst{1}
\and P.\,W.\,A.\,Roming\inst{5}
          }

  \offprints{D. Burrows}

\institute{
Department of Astronomy and Astrophysics,
The Pennsylvania State University,
525 Davey Lab,
University Park, PA  16802  USA
\and
\email{burrows@astro.psu.edu}
\and
Los Alamos National Laboratory,
Los Alamos, NM, USA
\and
Istituto Nazionale di Astrofisica --
IASF-Palermo, Palermo, Italy
\and
Southwest Research Institute, 
San Antonio, TX, USA
}

\authorrunning{Burrows et al.}

\titlerunning{Finding High-z GRBs}

\abstract{
We discuss the optimization of gamma-ray burst (GRB) detectors with a goal
of maximizing the detected number of bright high-redshift GRBs, in the
context of design studies conducted for the X-ray transient detector
on the \janus\ mission.  We conclude that the optimal energy band for
detection of high-$z$ GRBs is below about 30 keV.  We considered both
lobster-eye and coded aperture designs operating in this energy band.  Within the
available mass and power constraints, we found that the coded aperture
mask was preferred for the detection of high-$z$ bursts with bright
enough afterglows to probe galaxies in the era of the Cosmic Dawn.  This
initial conclusion was confirmed through detailed mission simulations
that found that the selected design (an X-ray Coded Aperture
Telescope) would detect four times as many bright, high-$z$ GRBs as
the lobster-eye design we considered.  The \janus\ XCAT instrument
will detect 48 GRBs with $z>5$ and fluence $S_x > 3 \times 10^{-7}$
erg cm$^{-2}$ in a two year mission.
}

\maketitle{}

\section{Introduction}

One of the most compelling problems facing astrophysics in the early
21$^{st}$ century is to understand the early Universe, and
in particular, the epoch known as the
Cosmic Dawn \citep{Astro2010}, when starlight from
the first generations of stars reionized the neutral intergalactic
medium, and when the structure of the Universe was set into the general
patterns that still prevail today.
Recent studies have shown that this reionization is complete
around redshift $z \sim 6$ \citep{Fan06}.  
The Universe at such high redshifts is challenging to study, but progress has been
rapid in recent years, with redshift records being broken repeatedly
through a variety of techniques \citep[e.g.][]{Tanvir09,Bouwens10a,Cucciara11}
involving observations of high-$z$ quasars, galaxies, and gamma-ray
bursts (GRBs).

Each of these probes of the high-$z$ universe has strengths and weaknesses.
Quasars are extremely luminous and can be bright sources even at 
high redshift, but the high UV and X-ray flux from the central engine
strongly affects the properties of the host galaxy, making it atypical.  Furthermore,
the density of quasars drops rapidly at $z>6$.
Normal
galaxies are being found, possibly at very high redshifts, in the Hubble Ultra
Deep Field \citep[e.g.][]{Bouwens10a} through photometric dropouts.  These
photometric redshifts are not entirely reliable, as the dropouts can also be
produced by dust extinction.  More importantly, the galaxies being
found represent the bright end of the luminosity distribution, and
even so, are so faint that the spectroscopy needed to probe their gas
is beyond the capabilities of current instrumentation.
GRBs, on the other hand, probe star formation regions in all types of
galaxies, and are so intrinsically bright that they can be used to
probe the interstellar medium (gas and dust content) of host galaxies
too faint to see, while the duration of the burst is short enough so
that only gas in the immediate vicinity of the central engine is
disturbed by the strong UV and X-ray emission.  However, the
short duration means that observations of GRBs must be carried out
rapidly, before the GRB fades too much.

\begin{figure}[tbp]
\resizebox{\hsize}{!}{\includegraphics[clip=true]{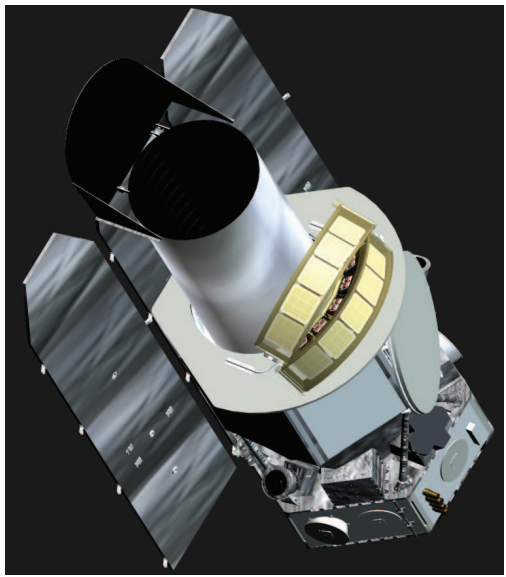}}
\caption{
\footnotesize
\janus\ design concept.  The mission features an X-ray Coded Aperture
Telescope (XCAT; gold-colored array of 10 modules mounted on the
satellite top deck) to find GRBs and X-ray transients, and a Near
Infrared Telescope (NIRT) that measures their redshifts with a
low-resolution objective prism spectrometer.
}
\label{fig:JANUS}
\end{figure}

The use of GRBs as probes of distant galaxies has blossomed in the
past six years \cite[e.g.][]{Prochaska06,Vreeswijk07,Prochaska07,Tumlinson07,Tejos08,Prochaska08,Chen09,DElia09,Levesque10b,Levesque10a}, 
since the launch of the \swift\ satellite
\citep{Gehrels04_Swift}, which provides rapid arcsecond positions of GRBs.
Still, after six years of operations, \swift\ has found only 3 GRBs
with spectroscopic redshifts exceeding 6 (Table~\ref{tbl:GRBs}).
\begin{table*}
\caption{Time delays to obtain redshift measurements for high-$z$ \swift\ GRBs}
\label{tbl:GRBs}
\begin{center}
\begin{tabular}{lccccl}
\hline
\\
GRB & $z$ & Type$^a$ & $T_P$ & $T_S$ & References \\
\hline
\\
050904 & 6.295 & S & 10 hr & 3.5 days & \citet{Cusumano06_GRB050904,
  Tagliaferri05}, \\
             &           &           &         &      & \qquad \citet{Kawai06} \\
060116 & 6.6 & P & 41 hr & N/A & GCN Circ. 4545 \\
080913 & 6.695 & S & 10 hr & 11 hr & \citet{Greiner09} \\
090423 & 8.2 & S & 7 hr & 24 hr &  \citet{Tanvir09} \\
090429B & 9.4 & P & 2.5 hr & N/A & \cite{Cucciara11} \\
\hline \\
\multicolumn{6}{l}{$^a$P = photometric redshift, S = spectroscopic
  redshift} \\
\multicolumn{6}{l}{$T_P$ = delay time to obtain photometric redshift;
  $T_S$ = delay time to obtain spectroscopic redshift}
\end{tabular}
\end{center}
\end{table*}
The paucity of high-$z$ bursts found to date probably stems from
several factors, 
one of which is almost certainly the long
delay in obtaining the first indication of a high-$z$ event, which is
typically many hours (Table~\ref{tbl:GRBs}).  The key to finding more high-$z$ bursts is to
design a satellite optimized to detect high-$z$ GRBs and to measure
their redshifts within minutes.  Here we discuss the optimization of
an instrument, the X-ray Coded Aperture Telescope (XCAT), designed to
find bright high-$z$ GRBs.

\section{The X-ray Coded Aperture Telescope on the Joint Astrophysics Nascent Universe Satellite}

\begin{figure}[tbp]
\resizebox{\hsize}{!}{\includegraphics[angle=90,bb=10 100 495 670, clip=true]{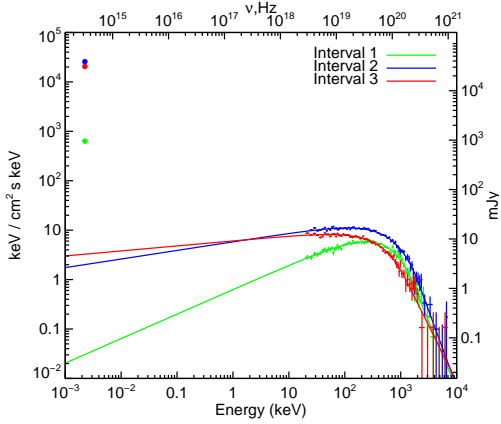}}
\caption{
\footnotesize
Prompt emission spectra for GRB\,080319B ($z =0.937$) for three time
intervals, plotted as $F_\nu$ in units of (keV cm$^{-2}$ s$^{-1}$ kev$^{-1}$).
Although there is strong spectral evolution, the spectra from all
three intervals are well-fit by a Band function (solid curves) with peak energy of a
few hundred keV.  From \citet{Racusin08}.
}
\label{fig:Band}
\end{figure}

\begin{figure}[tbp]
\resizebox{\hsize}{!}{\includegraphics[clip=true]{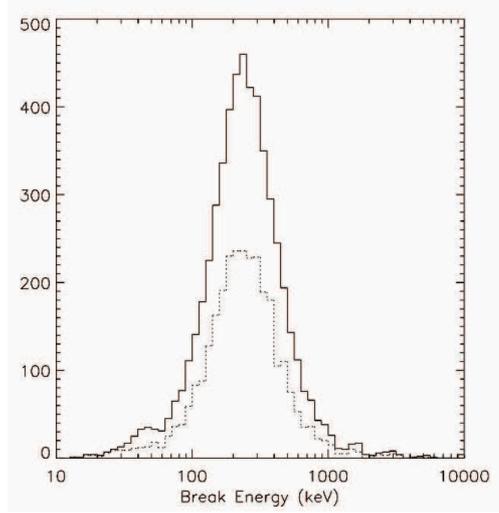}}
\caption{
\footnotesize
Distribution of peak energies for BATSE bursts, from \citet{Preece00}.  
}
\label{fig:BATSE}
\end{figure}
 
The Joint Astrophysics Nascent Universe Satellite
\citep[JANUS;][see Fig.~\ref{fig:JANUS}]{Burrows10_JANUS} is a proposed NASA Explorer mission
optimized to study the Cosmic Dawn by finding high-$z$ GRBs and
quasars and measuring their redshifts within minutes on-board.
It has two main instruments: an X-ray Coded Aperture Telescope (XCAT) \citep{Falcone10_XCAT}
and a Near Infrared Telescope (NIRT).  Here we discuss results of a
trade study made for the XCAT instrument to maximize the number of
bright high-$z$ bursts that we can discover.

Traditionally, GRB detectors have operated in the hard
X-ray to gamma-ray energy range, though {\it HETE-2} and {\it
  Beppo-SAX} both had soft X-ray capabilities.  From previous studies,
we know that the prompt emission from GRBs is typically characterized
as a ``Band function'' \citep{Band93}, an empirical function with a low
energy power law that transitions smoothly to a high energy power law,
with a peak energy ($E_p$, the peak in the $\nu F_\nu$ spectrum) 
near the transition point.  An example of time-resolved spectra of the
prompt emission from a bright burst at $z \sim 1$ is
shown in Figure~\ref{fig:Band}.   The distribution of BATSE peak 
energies is shown in Figure~\ref{fig:BATSE}: roughly 60\% of the bursts
have 100~keV $<E_p<$ 600~keV, with the most probable value at $E_p \sim 220$~keV.

 At high redshifts, the GRB prompt spectrum is shifted down in energy
by a factor of $(1+z)$.  We can therefore expect that high redshift
bursts will have peak energies in the tens of keV rather than the hundreds
of keV.  Below $E_p$ the photon spectrum is typically $I(E) \propto E^{-1}$
(photon index of $\Gamma \sim -1$). 
Above the peak energy, the photon spectrum drops off rapidly with energy
(typically as $E^{-4.5}$), and observations at energies above $E_p$
have poor sensitivity.  
Searches for high $z$ bursts should therefore concentrate on 
the region below about 30~keV, where
the number of photons per unit energy interval is high.
Because the spectrum favors the soft X-ray bandpass, and because soft
X-rays are much easier to detect than hard X-rays or $\gamma$-rays, a
small, lightweight soft X-ray instrument can have better
sensitivity to high-$z$ GRBs than a much larger, heavier hard X-ray or $\gamma$-ray
instrument, and is better suited to small missions in the SMEX or EX class.  

\subsection{Wide-Field X-Ray Imaging}

\begin{figure}[tbp]
\resizebox{\hsize}{!}{\includegraphics[clip=true]{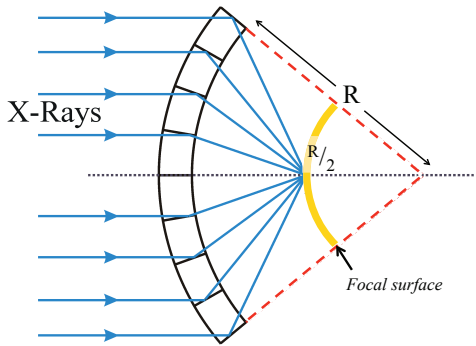}}
\caption{
\footnotesize
Cartoon of lobster-eye optics.  X-rays are reflecting at grazing
incidence from planar surfaces arranged in a spherical pattern (a
microchannel plate optic in this case).  The
thick yellow line represents the focal surface.
}
\label{fig:lobster_optics}
\end{figure}

Techniques available for performing wide angle X-ray transient surveys are severely restricted.
Large fields of view can be monitored with scanning collimator
instruments (such as the {\it Uhuru} satellite or the {\it MAXI}
mission), but these typically have poor position determination
accuracy (of order tens of arcminutes) that is insufficient to
localize GRBs and other transients well enough to identify
optical counterparts.  They also typically miss short transients like
most GRBs.
There are no materials with sufficient transparency and refractive power available to make
conventional lenses such as those used in optical cameras.
Large fields of view are feasible using multilayer-coated mirrors at
normal incidence \citep[e.g.][]{Smith90}, which work by
constructive interference of X-rays from layers of alternating
high-and-low Z materials, but these mirrors have very narrow bandpasses and
cannot be used for broad-band studies.
Broad-band X-rays can only be reflected at grazing incidence, making the
construction of conventional wide field optics virtually impossible. 
The relatively large grazing angles needed for wide fields of view
also limit the bandpass, since the maximum grazing angle is inversely
proportional to the photon energy.  The standard astronomical X-ray
telescope uses a Wolter I design \citep{Wolter52b,Wolter52a} that 
typically has fields of view less than a degree in diameter, with
strong vignetting near the edges of the field of view.

\begin{figure}[tbp]
\resizebox{\hsize}{!}{\includegraphics[clip=true]{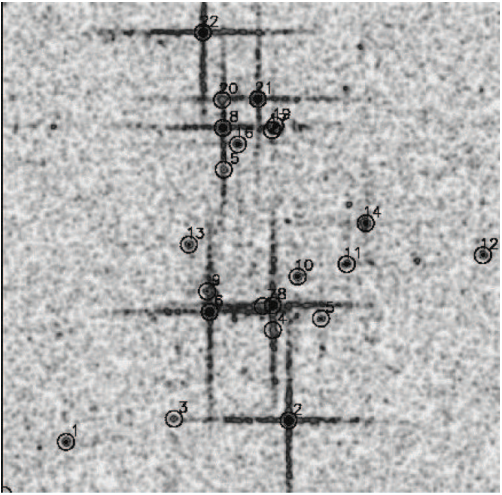}}
\caption{
\footnotesize
Simulation of an observation of a $10^\circ \times 10^\circ$
region of the LMC with the {\it LOBSTER-ISS} instrument (1 day exposure).  
From \citet{Fraser02}.  The square pore optics produce a cruciform PSF.
}
\label{fig:lobster}
\end{figure}

\begin{figure*}[t!]
\resizebox{\hsize}{!}{
\includegraphics[width=3.2in,clip=true]{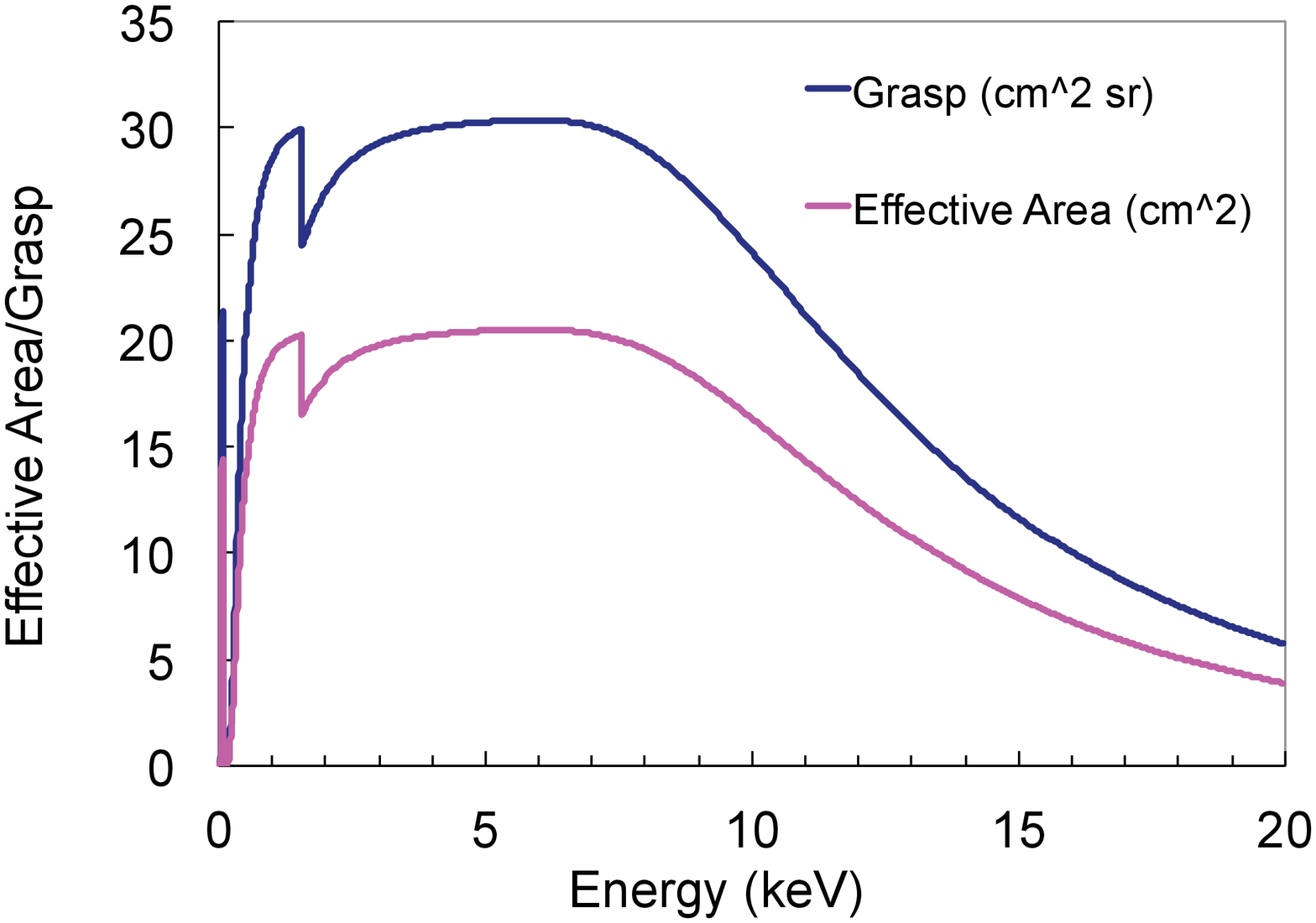}
\hfill
\includegraphics[width=2.8in,clip=true]{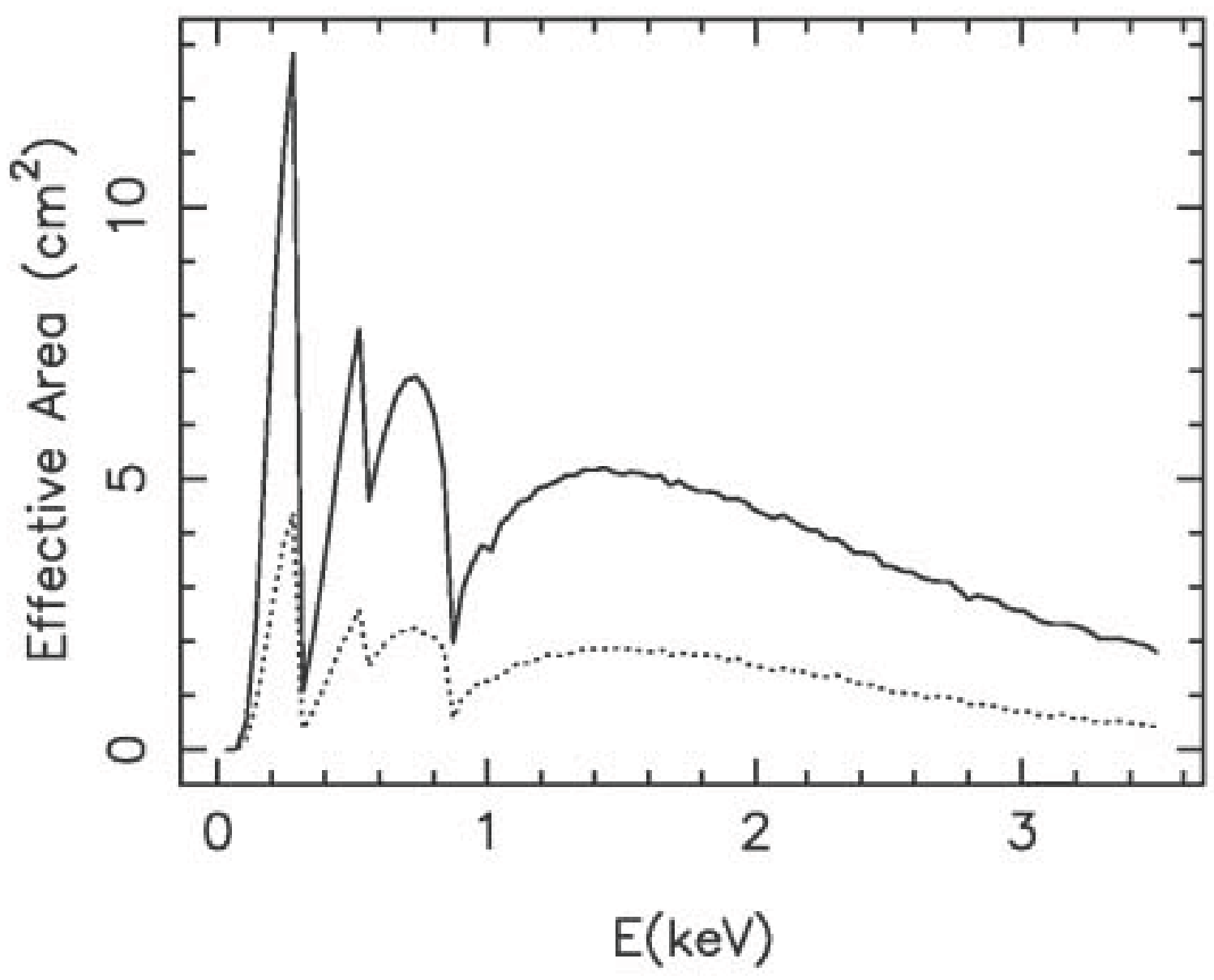}
}
\caption{\footnotesize
Left: XCAT grasp and effective area curves for a single module.  Right: LOBSTER-ISS
effective area curve \citep[from][]{Fraser02}.  The dotted curve shows
the effective area for the central true focus, while the solid curve
includes area scattered into the cruciform arms.
The XCAT effective area curve is much broader and
simpler because the coded mask cells are either transparent or opaque
over most of the energy range of the detector, so the curve is
dominated by the detector quantum efficiency.  The LOBSTER-ISS curve has
strong features caused by the grazing incidence reflections off of the
Ni-coated optic, and has a much narrower bandpass.
}
\label{fig:XCAT_EA}
\end{figure*}

These problems can be solved through the use of lobster-eye optics,
using a curved grazing incidence optic in which each segment
concentrates light from a different direction on the sky, but all reflections
are at grazing incidence (Figure~\ref{fig:lobster_optics}).  Such designs have been under development
for more than a decade, utilizing micropore optics or other systems of
reflectors \citep[e.g.][]{Angel79,Fraser02,Pearson03,Sveda09,Tichy09,Putkunz09}.
Lobster-eye designs have several disadvantages, including a complex Point
Spread Function (PSF; see Figure~\ref{fig:lobster} or
\citet{Hudec04}), 
small bandpass due to the reflective optics,
a complex energy-dependent effective area curve (e.g.,
Figure~\ref{fig:XCAT_EA}), and relatively
heavy optics.  Nevertheless, the sensitivity to point sources is
superior to coded apertures (discussed below).
Lobster-eye designs, however, have limited flight heritage.

For continuous, wide-field-of-view,
broadband, X-ray imaging with arcminute position determination, 
only coded aperture imaging has
significant flight history.  Since 1972, approximately two dozen
imaging coded aperture astronomical telescope designs
have flown in space or on balloons (see 
http://astrophysics.gsfc.nasa.gov/cai). 
Most instruments designed to detect GRBs have used 1-D or 2-D coded
aperture masks, using techniques developed and discussed
by a number of authors including \citet{Dicke68,Fenimore78,Proctor79}
and \cite{Caroli87}.
\begin{figure}[tbp]
\resizebox{\hsize}{!}{\includegraphics[clip=true]{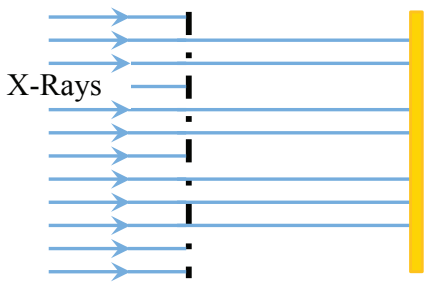}}
\caption{
\footnotesize
Cartoon of coded aperture ``optics''.  X-rays are either transmitted
or absorbed by cells in the mask (black) before impinging on the
detector plane (yellow).  We use a 2-D random mask pattern to
suppress ghost images.
}
\label{fig:coded_mask}
\end{figure}
Coded aperture masks work as multiple pinhole cameras or shadow
masks (Figure~\ref{fig:coded_mask}).  Each point source in the field of view casts a shadow of the
mask pattern onto the detectors.  As the angle of the source varies on
the sky, the shadow on the detectors is offset accordingly.  By
deconvolving the detector event pattern with the mask pattern, an
image of the sky can be obtained.  Two dimensional coded apertures
provide more robust positions on the sky than perpendicular 1-D coded
apertures (which can sometimes provide only a 1-D position and which
sometimes have ghost images).
The 2D coded mask, combined with sophisticated triggering algorithms,
gives the \swift\ BAT instrument an extremely low false burst rate of just a few percent.
Coded aperture instruments are in general less sensitive than
lobster-eye instruments because the photons are not concentrated into
images, and because every source and all diffuse emission contribute
noise to every pixel on the detector.  Nevertheless, coded aperture instruments have
advantages in terms of broad-band performance and extremely wide
fields of view for a given mass.

Three designs were considered in detail in our trade studies for \janus: a
lobster-eye design 
similar to the {\it LOBSTER-ISS} instrument proposed a decade ago for flight
on the International Space Station \citep{Fraser02}; a monolithic
coded aperture design similar to the Swift BAT, with a single large mask imaging onto a single
focal plane detector array; and a modular coded aperture design, in which each module pointed
in a different direction (X-ray Coded Aperture Telescope, or XCAT).
We found that sensitivity limits for monolithic and modular coded
aperture instruments are similar for on-axis sources, but that the
modular design performs better at the edges of the field of view (less
vignetting) and provided more design flexibility for accommodating
multiple instruments on a small platform.
We then compared the lobster-eye concept to the modular coded
aperture. 
Instrument parameters for the two design concepts we considered are given in
Table~\ref{tbl:comp}, and effective area curves are shown in 
Figure~\ref{fig:XCAT_EA}.  
Instrument mass and volume constraints limited the size of both
instruments; the lobster-eye design considered here is larger (larger
focal length) and more massive than XCAT, and covers only 7\% of
the solid angle of the XCAT instrument, but is much more sensitive.

\begin{table*}
\caption{Comparison of coded aperture and lobster-eye designs}
\label{tbl:comp}
\begin{center}
\begin{tabular}{lll}
\hline
\\
Instrument & XCAT (coded aperture) & Lobster-eye \\
\hline \\
Energy Range (keV) & 0.5--20 & 0.3--3 \\
Imaging technique & 2-D coded mask & Lobster-eye grazing incidence \\
FOV (sr) & 3.9 & 0.26 \\
Mask/Optics area (cm$^2$) & 1690 & 1600 \\
Focal length & 158 mm & $>350$ mm \\
Angular resolution (arcmin) & 6.3 & 4 \\
Focal Plane Area (cm$^2$) & 147 & 400 \\
Detector type & H2RG HyViSI CMOS & GEM \\
Power & 70 W & 51 W \\
Mass & 57 kg & 95 kg \\
Limiting sensitivity in \\
\qquad 30 s (mCrabs) & 240 (6.5$\sigma$) & 2.7 (5$\sigma$) \\
\# GRBs in 2 years with $5<z<12$ \\
\qquad and $S_x > 3 \times 10^{-7}$ erg cm$^{-2}$ & 48 & 11 \\
\hline
\end{tabular}
\end{center}
\end{table*} 

\section{Simulations}

\subsection{JANUS Mission Simulation}

Our detailed simulation began with a simulation of the mission
geometry for \janus.
The \janus\ mission will spend most of its time surveying nearly 1/2
of the high-latitude sky.  The \janus\ team produced a list of fields covering the
northern and southern galactic caps, and developed an algorithm to
select fields for observation.  A detailed mission simulation was
constructed that included realistic orbital parameters and spacecraft
design parameters, and calculated the position of the spacecraft every
minute for a two year mission.  At each one minute time step, the
spacecraft field of view was compared against observing constraints
(keep-out zones around the Sun, Moon, and Earth).  When one of these
constraints impinged on the field of view, a new target was selected
that would be observable for at least ten minutes, and the spacecraft
slewed to the new target (using realistic slew times based on a
detailed model of the spacecraft's attitude control system).  The
simulation correctly accounted for time spent in the South Atlantic
Anomaly (where the instruments cannot observe due to high background
rates), and included interruptions to the mission timeline for GRBs
that were injected at random times using the expected GRB event rate.
The mission simulation demonstrated that \janus\ can complete its
quasar survey in a two year mission with considerable margin, and
served as the basis for our detailed simulation of the XCAT instrument performance.

\subsection{XCAT GRB Simulation} 

To predict
\janus\ burst rates as a function of redshift, we
carried out a rigorous simulation of XCAT performance
on-orbit, making use of
the \janus\ mission simulation to
provide a realistic history of live-time periods
and satellite pointings (including relative Earth
position, to account for partial occultation of
the XCAT FOV). To generate an appropriate
population of GRBs for detection, we used the
best-fit luminosity and redshift distributions
from \citet{Wanderman10}, as updated
on the web to reflect the latest Swift results. We
drew burst redshifts ($0<z<30$) and luminosities
($L_{peak} \ge 10^{50}$~erg~s$^{-1}$) from these distributions
at Poissonian intervals according to the known
all-sky rate, placed the burst at a random position
on the sky, and compared to the FOVs of
the 10 XCAT modules (from the mission simulation). If the burst was in view
of one or more modules, then a simulated burst
light curve (with Poisson noise) was generated
for each viewing module -- including contributions
from all point X-ray sources in the module's
FOV along with the known diffuse X-ray
and particle backgrounds -- and was fed to the
XCAT triggering software. If the burst resulted in a trigger,
the properties of the burst were recorded.  These successful triggers
formed the basis for our predicted redshift distribution.
To minimize Poisson uncertainties due
to the Monte Carlo nature of the simulation,
we carried out 20 runs of 400 days (on-orbit)
duration each.

\begin{table*}
\caption{Redshift Distribution of Simulated \janus\ GRBs (2 year mission)}
\label{tbl:zdist}
\begin{center}
\begin{tabular}{ccccccccc}
\hline \\
$z$ & 5 & 6 & 7 & 8 & 9 & 10 & 11 & 12 \\
$f(>z)$ & 12\% &  6.4\% & 3.4\% & 1.8\% & 0.9\% & 0.4\% & 0.1\% & 0.16\% \\
$N(>z)$ & 61 & 33 & 17 & 9 & 5 & 2 & 1 & 1 \\
\hline
\end{tabular}
\end{center}
\end{table*}

Burst light curves were simulated by rescaling
the Swift BAT (15--150 keV) light curves of
bursts with known redshifts to the (new and
distinct) redshift and peak luminosity of the
simulated burst, and extrapolating
the observed burst spectrum to the lower energy
range of XCAT. (For simulated high-$z$ bursts
this extrapolation is minimal, since the median
redshift for the Swift burst sample is only $z=2.3$).
We have developed a library of 111 Swift BAT
light curves for this purpose.

\begin{figure}[tbp]
\resizebox{\hsize}{!}{\includegraphics[clip=true]{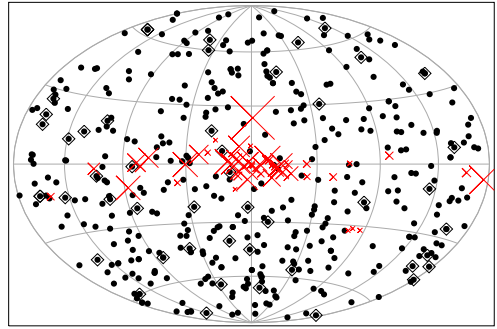}}
\caption{
\footnotesize
Distribution of simulated GRBs on the sky for the XCAT instrument
design (2 year mission).  
Black diamonds indicate the high-$z$ GRBs.
Red crosses show the location of the 50
brightest X-ray
sources seen by the {\it RXTE} All-Sky Monitor (ASM); 
the 263 sources detected by the ASM during the week of 7
February 2009 were included as background sources in the
simulation.   (Cross size indicates relative source brightness.)
}
\label{fig:GRB_map}
\end{figure}

\begin{figure}[tbp]
\resizebox{\hsize}{!}{\includegraphics[clip=true]{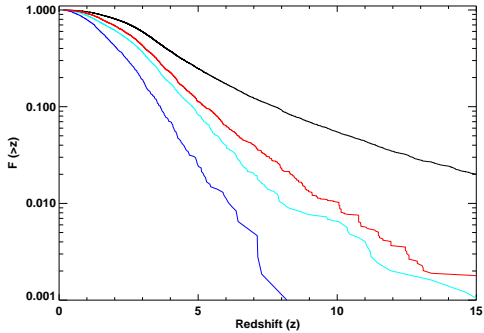}}
\caption{
\footnotesize
Redshift distribution of GRBs.  The black line shows the distribution
of the parent population from \citet{Wanderman10}.  The red line shows
the results of our simulation for the XCAT instrument.  The cyan line
shows the distribution for the \swift\ BAT instrument.  The dark blue
line shows the distribution for BATSE.  We expect XCAT to detect
a significantly higher fraction of high-$z$ GRBs than \swift; this,
combined with rapid redshift measurements for every high-$z$ GRB
detected by XCAT, will result in about 30 GRBs per year with $z>5$.
}
\label{fig:zdist}
\end{figure}

On the basis of the simulations, we predict a
mean burst detection rate of 0.7 GRBs per day
with XCAT.  
The distribution of these bursts on the sky is shown in Figure~\ref{fig:GRB_map}.
The redshift distribution of detected bursts is given in
Table~\ref{tbl:zdist} and is shown in Figure~\ref{fig:zdist}.
Beyond $z=12$ the Ly$\alpha$ break
redshifts beyond the NIRT bandpass so that a
redshift measurement is not possible. 
The XCAT design will detect 61 bursts with $z>5$ in a 2 year mission.

\subsection{Lobster-Eye Simulation}

\begin{figure}[tbp]
\resizebox{\hsize}{!}{\includegraphics[clip=true]{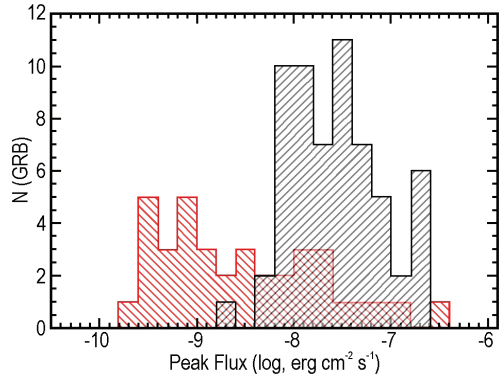}}
\caption{
\footnotesize
Distribution of peak fluxes of high-redshift GRBs ($5 < z < 12$) 
detected by XCAT (black) and the lobster-eye design considered here
(red).  The two instruments detect similar numbers of high-$z$ GRBs, but the
XCAT distribution is much brighter.
}
\label{fig:flux_distr}
\end{figure}

The GRBs generated during the detailed XCAT simulation were also used
as input to a simulation of the lobster-eye design.  The
instrument parameters used for the simulation are shown in
Table~\ref{tbl:comp}.
We found that the lobster-eye instrument detected about the same total number of bursts
in spite of its small field of view, because the improved sensitivity
allowed it to detect every GRB that went off in its field of view.
Although the total number of bursts is similar, the distribution of
peak flux is very different from that of the XCAT instrument
(Figure~\ref{fig:flux_distr}).

\subsection{Optical / Gamma-ray Correlation}

This difference in peak flux distributions of the high redshift burst
samples detected by XCAT and by our lobster-eye implementation 
is quite significant.  Previous
studies have shown that the brightness of the optical afterglow is
correlated with the brightness of the prompt $\gamma$-ray emission
\citep{Gehrels08,Nysewander09,Kann10}.  Although there is considerable
scatter (perhaps due to dust extinction affecting the optical
brightness), the trend is that brighter GRBs (measured by peak flux,
total fluence, or total isotropic energy output) have brighter optical
afterglows. 
Bright afterglows are essential for detailed high resolution studies of
the host galaxies of the bursts.
 {\it Therefore, in order to use high redshift bursts to
probe galaxies at the Cosmic Dawn, one should optimize the burst
detection to maximize the number of bright  bursts.}
We find that the XCAT design will detect 24 high-$z$ GRBs per year at
fluence $S_x > 3 \times 10^{-7}$ erg cm$^{-2}$, while the lobster-eye design
detects only 6 per year.

\section{Conclusions}

Our detailed simulations confirmed our ``back-of-the-envelope''
determination that high-$z$ GRBs are best detected in the soft X-ray
band, by which we mean energies of roughly 0.5-20 keV.  In this energy
band, the best options for wide-field transient detectors are
lobster-eye and coded aperture designs.
The lobster eye
design is more sensitive and will find more faint X-ray transients,
but its field of view is too small to find many rare, bright, high-$z$
GRBs that can be used to probe galaxies in the era of the Cosmic Dawn
with high resolution spectroscopy followup.  The XCAT coded mask
design was selected for the \janus\ mission because it is optimized to
find these high-$z$ bursts, and therefore provides the best solution
to the prime \janus\ science goals of probing the Cosmic Dawn with
GRBs and quasars.
 

\bibliographystyle{aa}
\bibliography{references}

\end{document}

%% file: definitions.tex
\newcommand{\cgs}{\mbox{${\rm ergs~cm}^{-2}~\rm{s}^{-1}$}}
\newcommand{\chisq}{\mbox{$\chi^2$ }}	    
\newcommand{\chinu}{\mbox{$\chi_{\nu}^2$}}  
\newcommand{\cmcm}{\mbox{${\rm cm}^{-2}$}}
\newcommand{\degrees}{\mbox{$^{\circ}$}}    
\newcommand{\degdot}{\mbox{$. \! ^{\circ}$}} 
\newcommand{\Ha}{\mbox{${\rm H {\alpha}}$}}  
\newcommand{\HI}{\mbox{H\scriptsize I}}
\newcommand{\HII}{\mbox{H\scriptsize II}}
\newcommand{\ir}{\mbox{$100 \mu$m}}	      
\newcommand{\kms}{\mbox{km ${\rm s}^{-1}$}}
\newcommand{\NH}{\mbox{${\rm N}_{\rm H}$}} 
\newcommand{\NHunits}{\mbox{$\times 10^{20} {\rm cm}^{-2}$}}
\newcommand{\NII}{\mbox{N\scriptsize II}}
\newcommand{\persec}{\mbox{${\rm s}^{-1}$}}    
\newcommand{\oq}{\mbox{$\frac{1}{4}$}}	    
\newcommand{\tq}{\mbox{$\frac{3}{4}$}}	    

\newcommand{\meszaros}{M\'{e}sz\'{a}ros}
\newcommand{\Meszaros}{M\'{e}sz\'{a}ros}
\newcommand{\peterm}{M\'{e}sz\'{a}ros}
\newcommand{\bohdan}{Paczy\'{n}ski}
\newcommand{\Bohdan}{Paczy\'{n}ski}
\newcommand{\paczynski}{Paczy\'{n}ski}
\newcommand{\Paczynski}{Paczy\'{n}ski}

\newcommand{\ApJ}{{\it Ap. J.}} 	    
\newcommand{\ApJL}{{\it Ap. J. (Letters)}}  
\newcommand{\AJ}{{\it A. J.}}		    

\newcommand{\etal}{{et al. }}		

\newcommand{\alexis}{{\it ALEXIS}}
\newcommand{\chandra}{\mbox{\it Chandra}}	    
\newcommand{\cobe}{{\it COBE}}
\newcommand{\heao}{{\it HEAO-1}}
\newcommand{\iras}{{\it IRAS}}
\newcommand{\rosat}{{\it ROSAT}}
\newcommand{\swift}{{\it Swift}}